\documentclass[apj]{emulateapj}
\begin{document}
\bibliographystyle{apj}

\title{Turbulent Amplification and Structure of Intracluster Magnetic Field}

\author{Andrey Beresnyak}
\affil{Nordita, KTH Royal Institute of Technology and Stockholm University, SE-10691 Stockholm, Sweden}
\author{Francesco Miniati}
\affil{Physics Dept., ETH Zurich, Wolfgang-Pauli-Strasse 27, CH-8093 Switzerland}

\begin{abstract}
We compare DNS calculations of homogeneous isotropic turbulence with the statistical properties of intra-cluster turbulence from the Matryoshka Run \citep{2014ApJ...782...21M} and find remarkable similarities between their inertial ranges. This allowed us to use the time dependent
statistical properties of intra-cluster turbulence to evaluate dynamo action in the intra-cluster medium, based on earlier results from numerically resolved 
nonlinear magneto-hydrodynamic turbulent dynamo \citep{B12a}. We argue that 
this approach is necessary (a) to properly normalize dynamo action to the available intra-cluster turbulent energy and
(b) to overcome the limitations of low {\it Re} affecting current numerical models of the intra-cluster medium.
We find that while the properties of intra-cluster magnetic field are largely insensitive to the value and origin 
of the seed field, the resulting values for the Alfv\'en speed and the outer scale of the
magnetic field are consistent with current observational estimates, basically confirming
the idea that magnetic field in today's galaxy clusters is a record of its past turbulent activity.  
\end{abstract}


\section{Introduction}
The hot intracluster medium (ICM) of galaxy clusters (GC) is well known to
be magnetized from radio observations.  These reveal both the
occurrence of Faraday rotation effect on polarized radiation from
background quasars~\citep{2001ApJ...547L.111C,2004JKAS...37..337C} and
of diffuse synchrotron emission~\citep{2008SSRv..134...93F} from the
ICM. Estimates of the magnetic field based on these observations
range between a fraction and several $\mu$G. Measurements on 
the structural and spectral features are sparse and more difficult, 
but indicate steep power-laws below few tens of kpc~\citep{2008MNRAS.391..521L,2011A&A...529A..13K}.
For massive clusters, turbulence in the ICM is mainly driven by structure formation~\citep{1999LNP...530..106N,2008Sci...320..909R,2011A&A...529A..17V,2014ApJ...782...21M,2015ApJ...800...60M}.
The most important magnetic field amplification mechanism in the ICM is the small scale or fluctuation dynamo (SSD), operating on scales smaller than the turbulence outer scale. Kinematic regime of SSD, i.e. when 
the back reaction of the magnetic field on the flow is negligible, has been studied in great detail previously~\citep{kazantsev1968,kraichnan1967,kulsrud1992}.
In kinematic regime the magnetic energy grows exponentially, till the approximation
breaks down, roughly in a dynamical time multiplied by $Re^{-1/2}$, where $Re$ is an effective
Reynolds number. 
The extremely hot and rarefied plasma of the cluster have very large collisional
mean free paths, around 
\begin{equation}
\lambda\approx 10^3\, {\rm pc} (n/3\times10^{-3} {\rm cm}^{-3})^{-1}  (T/10 {\rm keV})^{3/2}, 
\end{equation}
at the same time, given the observable magnetic fields around 3 $\mu$G, 
the Larmor radius is smaller by many orders of magnitudes:
\begin{equation}
r_L\approx 10^{-9}\, {\rm pc} (T/ 10{\rm keV}) (B/3 \mu G)^{-1}.
\end{equation}
Such situation, known as ``collisionless plasma'' is challenging from theoretical viewpoint, since nonlinear plasma effects are dominating the transport, which has been
known since early Lab plasma experiments, when it became clear that collisional ``classic transport'' is grossly insufficient to explain cross field diffusion \citep[see, e.g.,][]{galeev1979}. As a rule of thumb, the actual effective parallel mean free path is smaller
than the one obtained by collisional formula, but larger
than the Bohm estimate ($\lambda_{eff} \sim r_L$). The search for this ``mesoscale'' for cluster conditions resulted in estimates for the mean free path of the proton in the ICM around $10^{-3}-10^{-6}$pc \citep{Schekochihin2006,BL06,Schekochihin2008,brunetti2011b}.
From these estimates we expect clusters to be turbulent with Reynolds numbers Re exceeding $10^{12}$. Combining this with the above estimate of the kinematic SSD growth
rates, for a dynamical time $\sim$ eddy turnover time $\sim$ 1 Gyr~\citep{2014ApJ...782...21M}, 
we estimate that the exponentiation timescale will be smaller than 1 Gyr $(Re)^{-1/2}\approx $ 1 kyr. 

The remainder of this paper is organized as follows: in Section~2 we
discuss the properties of nonlinear regime of the small-scale dynamo which is supposed
to dominate during most of the cluster lifetime; in Section~3 we point to the inadequacy
of current MHD cosmological simulations, as far as dynamo is concerned, and suggest
a different approach; in Section~4 we describe new homogeneous dynamo simulations with intermittent driving; in Section~5 we explain our cosmological hydrodynamic model of the cluster; in Section~6 we combine the knowledge obtained in previous sections and analyze cluster simulations to derive the properties of the cluster magnetic fields; in Section~7 we discuss implications and compare with previous work.

\section{Nonlinear Small-scale Dynamo}\label{SSD:sec}
As the kinematic approximation of SSD breaks down very quickly, 
the dynamo spends most of the time in the nonlinear regime.
In this regime, inclusive of the back reaction of the magnetic field on the flow,
the magnetic energy continues to grow as it reaches equipartition with the turbulent
kinetic energy cascade at progressively larger scales~\citep{schluter1950}.
At this stage the magnetic energy
is characterized by a steep spectrum and an outer scale, $L_B$, a small fraction
of the kinetic energy outer scale~\citep{haugen2004,brandenburg2005,ryu2008,CVB09}.
This picture has been later argued to be true in any high-$Re$ flow,
with the argument relying on locality of energy transfer functions~\citep{B12a}.
It also followed from this study that the growth rate of the magnetic energy corresponds to 
a certain fraction of the turbulent dissipation rate, with this fraction being
a universal dimensionless number around $0.05$, and tat
the magnetic outer scale $L_B$ grows with time as $L_B\approx t^{3/2}$~\citep{B12a}.
The growth of magnetic energy reaches final saturation when $L_B$ is a substantial 
fraction of the outer scale of the turbulence. However, this never happens in clusters, as we show below.

\section{Limitations of cosmological dynamo simulations}

\begin{figure}
\includegraphics[width=1.0\columnwidth]{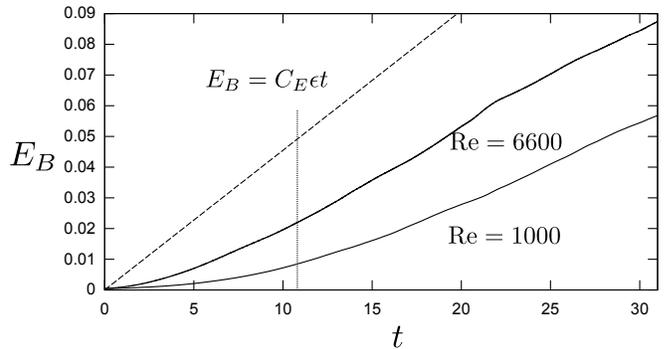}
\caption{The growth of dynamo can be significantly delayed in simulations
compared with actual high-Re flows, which is depicted here. Here we compare 
the growth of magnetic energy from $4\times 10^{-4}$ below equipartition
in $Re=10^3$ and $Re=6.6\times 10^3$ simulations (solid) from \cite{B12a} and the hypothetical Re=$\infty$ case which is represented by linear growth $C_E \epsilon t$ (dashed). If we stop simulation around the dashed line, which roughly correspond to the duration of cosmological simulation $\sim 11$ dynamical times, the magnetic energy
will be grossly underestimated and its value will depend on both the seed field
and the effective Reynolds number of the simulation.}
\label{delayed}
\end{figure}

An important implication of the above picture is that the memory of the initial
seed field is quickly lost and the cluster magnetic field is expected
to depend only on the cluster turbulent history.
While this theoretical insight was certainly useful, its applications
to cluster formation were not immediately realized.
There are two main reasons
for this. Firstly, while there has been considerable progress in
computational models of structure formation, and GCs in
particular, the level of dynamic range of spacial scales achieved so
far is considerably below the threshold necessary for the turbulent
dynamo to operate efficiently.  In fact, numerical MHD models of
GCs typically report rather weak magnetic field
amplification roughly by factors $\lesssim
30$~\citep{2001ApJ...562..233M,2002A&A...387..383D,2008A&A...482L..13D,2014MNRAS.445.3706V},
including significant contribution from adiabatic compression.  As alluded
above, the reason is ascribed to the low $Re$ of the simulated
flows. The kinematic growth rate is $\gamma\approx {Re}^{1/2}/30 \tau_L$
\citep{haugen2004,Schekochihin2004,B12a}, where $\tau_L$
is the turnover time of the largest eddy.
So even with $Re\sim$ several $\times 10^2$, typical for cluster
simulations, the dynamo will be stuck for several dynamical
times in a kinematic regime, i.e. several Gyr, while in nature this stage will be many
orders of magnitude quicker than the dynamical time (see also Section~1).
Fig.~\ref{delayed} demonstrates the difference between the magnetic energy growth between
the case with very large Re (straight line) and Re that are available with current
numerical capabilities (actual growth obtained in simulations with Re=1000 and 3300).
The growth observed in simulations is delayed due to the grossly prolonged kinematic stage. 
Secondly, in view of the current understanding of MHD dynamo (Section~\ref{SSD:sec}), 
lack of detailed knowledge about the ICM turbulence precludes accurate estimates of both 
the magnetic energy and, in particular, the outer scale of the magnetic spectrum.

Below we report on the progress with the approach which is different from direct approach of cosmological MHD simulation, which, given present state of our numerical capabilities, as we argued above 
is completely inadequate.
We have recently employed a novel technique to model the formation of a
massive GC with sufficient resolution to resolve the
turbulent cascade~\citep{2014ApJ...782...21M,2015ApJ...800...60M}.  We have extracted
the time dependent properties of the turbulence
and used this information in combination with independent results
on turbulent dynamo obtained from high resolution periodic box simulations.
The novelty and advantage of our approach is that the turbulence is self-consistently
estimated through a numerical hydrodynamic model of structure formation, while the magnetic field
evolution is estimated based on theory, which was confirmed in large-scale
homogeneous dynamo simulations, robustly tested by studying low Re effects
in a scaling study. Importantly enough, such dynamo simulations, unlike cosmological cluster
models, are not limited in the number of dynamical times one can simulate.

\section{Dynamo simulations}
\begin{figure}
\includegraphics[width=1.0\columnwidth]{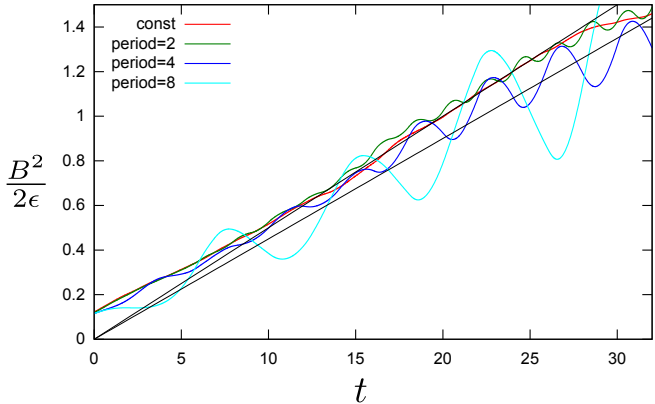}
\caption{We plotted the magnetic energy, $B^2/2$ (Alfv\'en units), divided by the energy driving rate, $\epsilon$, resulting in $B^2/2\epsilon$ (simulation time units) versus time (simulation time units), similar to results reported in \cite{B12a}. Several curves correspond to intermittent driving with the period of driving varied from 1 to 8 $\tau_c$ and the half of the period the driving was on while the other half off. We also put a simulation with constant driving for comparison. Two thin lines correspond to the conversion efficiencies of nonlinear dynamo $C_E$ of 0.045 and 0.05 (dimensionless).}
\label{b_field}
\end{figure}

\begin{figure}
\includegraphics[width=0.95\columnwidth]{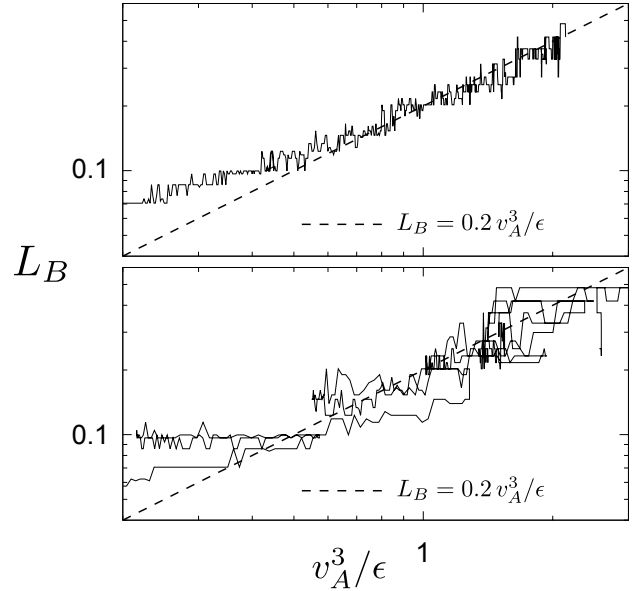}
\caption{The relation between magnetic energy and the outer scale in the DNS of nonlinear
small scale dynamo. The upper plot correspond to the case with constant driving, reported earlier in \citet{B12a}, while the lower plot has been produced in otherwise similar simulation with intermittent driving with period of $8 \tau_c$ (self-correlation timescales). In the case of
intermittent driving we used the dissipation rate $\epsilon$ averaged over $2 \tau_c$ in a manner similar to the analysis of cluster data. 
We defined outer scale of magnetic field through its peak wavenumber of the spectrum $L_B=2\pi/k_{\rm max}$. The best fit corresponds to the coefficient $c_l\approx0.18$ in Eq.~(\ref{Lb}).}
\label{eblb}
\end{figure}

We have extended the study of statistically homogeneous isotropic small-scale dynamo simulations in~\citet{B12a} with a series of simulations with intermittent energy injection into the velocity field, with the period 1,2,4, and 8 self-correlation timescales of velocity, $\tau_c$. All simulations have magnetic Prandtl number $Pr_m=1$ and
driving in Fourier space was limited to lower harmonics ($|k|<2.5$). We started each MHD simulation by seeding low level white noise magnetic field into the dataset obtained from driven hydrodynamic simulation which reached statistically stationary state. This
dataset was further evolved by full incompressible MHD equations. Figure~\ref{b_field} shows
the evolution of magnetic energy in time. The previously measured normalized growth rate $C_E=0.05$ is roughly consistent with most of the data. An important prediction of \citet{B12a} was also that the magnetic outer scale is proportional to $v_A^3/\epsilon$
and grows in time as $t^{3/2}$. Since we are going to use this conjecture to estimate
the outer scale of cluster magnetic fields, we plotted the $v_A^3/\epsilon$ versus the magnetic outer scale, which we determined from the peak of magnetic spectrum. The constant driving simulation (upper panel of Figure~\ref{eblb}) showed good agreement with the proposed scaling and we have determined the dimensionless coefficient $c_l$ in the relation $L_B=c_l v_A^3/\epsilon$ to be around 0.2 (best fit 0.18). The intermittently driven simulation
have shown large scatter which is due to the fact that turbulence spectra do not depend instantaneously on the energy injection rate, but have a memory of the previous state
around about one dynamical time. Also, the cascade rate at the equipartition scale
is delayed compared to the injection rate. We found that averaging cascade rate over
$2 \tau_c$ and introducing phase delay of $\pi/2$ will work the best to reproduce the $L_B$ relation and we presented the plot of such constructed $L_B$ - $v_A^3/\epsilon$
on the lower panel of Figure~\ref{eblb}. The scatter was significantly reduced and
the derived $c_l$ coefficient is also the same as for the constant driving case. 

\section{Cluster simulations} 
We use the Matryoshka run to extract the time dependent turbulence properties 
of the ICM of a massive GC, with total mass  at redshift $z=0$ of $1.3\times 10^{15}M_\odot $,
forming in a concordance $\Lambda$-CDM universe~\citep{2009ApJS..180..330K}.
The simulation was carried out with {\tt CHARM}, an
Adaptive-Mesh-Refinement cosmological code~\citep{2007JCoPh.227..400M}.
We use a concordance $\Lambda$-CDM universe with normalized (in units
of the critical value) total mass density, $\Omega_m=0.2792$, baryonic
mass density, $\Omega_b=0.0462$, vacuum energy density,
$\Omega_\Lambda= 1- \Omega_m= 0.7208$, normalized Hubble constant
$h\equiv H_0/100$ km s$^{-1}$ Mpc$^{-1}$ = 0.701, spectral index of
primordial perturbation, $n_s=0.96$, and rms linear density
fluctuation within a sphere with a comoving radius of 8 $\,h^{-1}$
Mpc, $\sigma_8=0.817$~\citep{2009ApJS..180..330K}.
The simulated volume has comoving size of $L_{Box}=240\,h^{-1}$ Mpc on
a side.  The initial conditions were generated on three refinement
levels with~\texttt{grafic++} (made publicly available by
D. Potter). For the coarsest level we use 512$^3$ comoving cells,
corresponding to a nominal spatial resolution of 468.75$\,h^{-1}$
comoving kpc and 512$^3$ particles of mass $6.7\times 10^9\,h^{-1}$
$M_\odot$ to represent the collisionless dark matter component.  The
additional levels allow for refined initial conditions in the volume
where the galaxy cluster forms.  The refinement ratio for both levels
is, $n_\mathrm{ref}^\ell\equiv \Delta x_{\ell}/\Delta x_{\ell+1}=2$,
$\ell=0,1$.  Each refined level covers 1/8 of the volume of the next
coarser level with a uniform grid of 512$^3$ comoving cells while the
dark matter is represented with 512$^3$ particles.  At the finest
level the spatial resolution is $\Delta x=$ 117.2$\,h^{-1}$ comoving
kpc and the particle mass is $10^8\,h^{-1}$ M$_\odot$.  As the
Lagrangian volume of the galaxy cluster shrinks under self-gravity,
three additional uniform grids covering 1/8 of the volume of the next
coarser level were employed with 512$^3$, 1024$^3$ and 1024$^3$
comoving cells, respectively, and $n_\mathrm{ref}^\ell=2,4,2$, for
$\ell=2,3,4$, respectively.  All of them were in place by redshift 1.4,
providing a spatial resolution of 7.3 h$^{-1}$ comoving kpc in a
region of 7.5 h$^{-1}$ Mpc, accommodating the whole virial volume of
the GC. The ensuing dynamic range of resolved spatial scales is
sufficiently large for the emergence of turbulence.
The results of the
cluster simulation is described in full detail in~\citep{2014ApJ...782...21M,2015ApJ...800...60M}.

\section{Analysis of cluster simulations} 
Using a Hodge-Helmholtz decomposition it was found that between 60 and 90\%
of the kinetic energy of the cluster turbulence is in the solenoidal
component~\citep[][see also~\citet{2011ApJ...731...62F}]{2015ApJ...800...60M}. This is the relevant component
for the discussed small-scale dynamo mechanism and the key question
is whether it resembles homogeneous isotropic turbulence in the inertial range. 

\begin{figure}
\includegraphics[width=0.95\columnwidth]{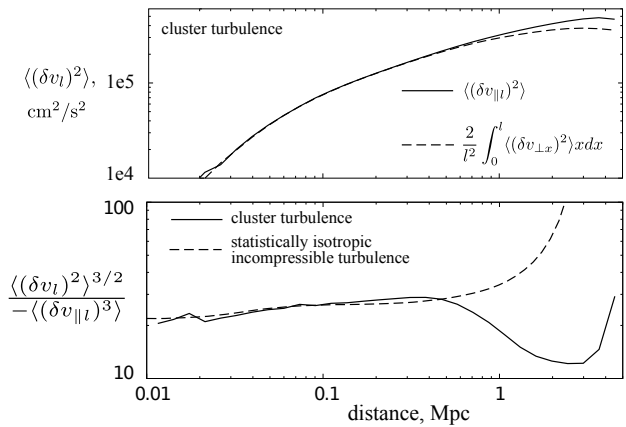}
\caption{We used structure functions to test statistical isotropy of
  the given cluster simulations (upper panel). Also we compare the
  relation between second order total structure function and the third order parallel structure
  functions in the cluster simulation (solid line) and in a periodic box simulation of statistically
  stationary driven turbulence (dashed line), after rescaling
  the dissipation scale to the same number (lower panel). Note
  that no fitting has been involved in making the both figures.
  We conclude that the second order structure function in the range
  0.14-0.4 Mpc could be used to robustly estimate the turbulent
  dissipation rate associated with the solenoidal velocity component.}
\label{hydro}
\end{figure}

In Fig~\ref{hydro}, upper panel we checked statistical isotropy
of the cluster turbulence. We compared the longitudinal velocity structure function (SF) with the analytical expression that presumes statistical isotropy.
Statistical isotropy seems to be satisfied quite well on all scales of interest
consistent with results in~\citep{2015ApJ...800...60M}.
A more critical test is provided by the relation
between structure functions of different order. For example the dimensionless ratio
$\langle(\delta v_l)^2\rangle^{3/2}/\langle(\delta v_{\|l})^3\rangle$ is of interest
to relate the energy cascade rate with the energy content of the cascade.
In the lower panel of Fig~\ref{hydro} we studied the comparison between this ratio
in the cluster simulation and in the homogeneous incompressible driven turbulence.
For the latter we used data from fully resolved direct numerical simulation
of incompressible hydrodynamic driven turbulence in a periodic box, see, e.g. \citet{BL09b}.
Both cluster and box simulation exhibited a clear well-pronounced dissipation interval
which we used to convert box simulation units into physical scale units of the cluster
simulation. Note, however, that no fitting has been involved on the y-axis.
Given relatively short inertial range the correspondence between homogeneous isotropic
turbulence statistics and cluster statistics is quite remarkable.
From the above comparison we conclude that the second order 
structure function of the cluster simulations in the range of scales
0.14-0.4 Mpc could be reliably used to estimate the turbulent
dissipation rate, $\epsilon_{turb}$, associated with the incompressible velocity component
and necessary to evaluate dynamo action in the ICM.

The turbulence dissipation rate is then estimated as follows:
\begin{equation}
\epsilon_{turb}=(c_1/c_2) (5/4) \langle(\delta v_l)^2\rangle^{3/2}/l,
\end{equation}
where $c_2\approx 27$ is the 
ratio of the structure functions reported in Fig.~\ref{hydro} and $c_1 \approx 1.17$
is a factor to correct for dissipation effects, as in our
finite Re simulations the Kolmogorov's -4/5 normalization 
slightly underestimates the turbulent dissipation rate. 

\begin{figure}
\includegraphics[width=0.95\columnwidth]{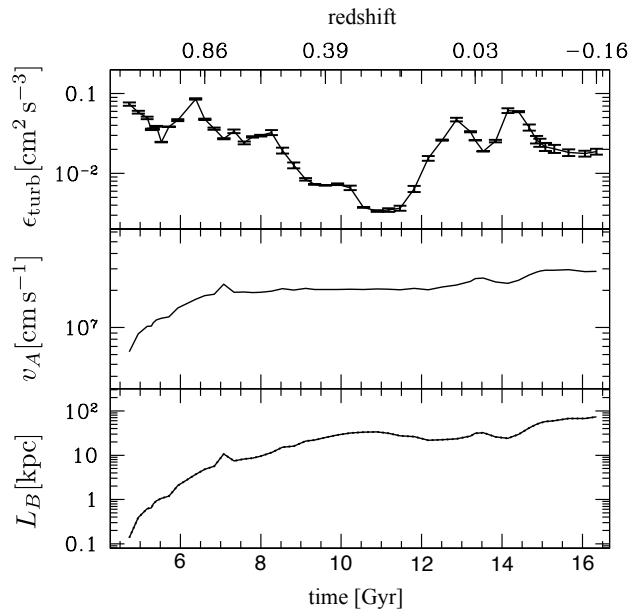}
\caption{The evolution of the turbulent dissipation rate (upper plot), Alfv\'en speed (middle) and the outer scale of the magnetic field (bottom) inferred from the cosmological cluster simulation by using
self-similar laws for turbulence and dynamo tested and normalized in well-resolved high-$Re$ DNS. In obtaining $v_A$, we divided the magnetic energy $E_B$ from Eq.~\ref{e_b}
to the average density within 1/3 of the virial radius.}
\label{cluster}
\end{figure}

As expected, at a given time $\epsilon_{turb}$ is a rather constant function of $l$ within the inertial range. 
The observed deviation was used to estimate the error of the measurement
of $\epsilon_{turb}$.  We plotted the dissipation rate determined
in this manner on the top panel of Fig~\ref{cluster}. We used the velocity structure
function calculated within 1/3 of the virial radius of the simulated cluster for each
data-cube. As we see from this figure, the
dissipation rate varies non monotonically over roughly an order of magnitude in
scale over the lifetime of the cluster. The errorbars defined above
indicate the deviation from Kolmogorov's self-similarity and were rather small,
except for the time intervals where the rate was changing rapidly,
i.e. the cluster was either relaxing of experiencing a fresh injection
of kinetic energy.

We then estimated magnetic energy density as~\citep{B12a}
\begin{equation}
E_B=\int_0^t C_E\rho\epsilon_{turb} dt
\label{e_b}
\end{equation}
with $C_E=0.05$. We plotted Alfv\'en
velocity on the middle panel of Fig~\ref{cluster}. Furthermore, as was
shown in \cite{B12a}, for statistically stationary turbulence the
magnetic energy containing scale could be estimated as
\begin{equation}
L_B=c_l v_A^3/\epsilon_{turb}, \label{Lb}
\end{equation}
where $c_l\approx 0.18$ is a universal coefficient, which could be determined in
DNS, see Fig.~\ref{eblb}.  
Our cluster turbulence was rather non-stationary, however,
as discussed in Section 4, the estimate Eq.~(\ref{Lb}) can also be applied to non-stationary driven turbulence
as long as the dissipation rate is averaged over a timescale around one dynamical time,
see below. This is because hydrodynamic cascade has a memory
over around one dynamical time and the changes in the driving rate do not instantaneously
affect turbulent rate on small scales (Section 4).
So, in using Eq.~(\ref{Lb}) we used the $\epsilon_{turb}$ averaged over 2 Gyr,
which approximately corresponds to two dynamical times.

The middle and bottom panels of Fig~\ref{cluster} show time evolution of the average
RMS Alfv\'en speed, $v_A=(2 E_B/\rho)^{1/2}$ and the magnetic outer scale $L_B$. Note that
while $v_A$ grows monotonically, $L_B$ can decrease somewhat during prolonged increase
of the turbulent activity, such as during several major mergers.

Our estimates for $z\sim 0$ characteristic values of
Alfv\'enic speed $v_A\sim 10^7 \rm cm/s$ and
the outer scales $L_B \sim 30-50$ kpc, are consistent with
the observed values reported in the literature
\citep{Mcnamara2007,Eilek2002,Bonafede2010,Govoni2006,Govoni2010}.
This indicates that the type of nonlinear dynamo described in
\cite{B12a} is probably operating in clusters, while kinematic models
would be challenged to achieve this.

One interesting conclusion from our results on Fig~\ref{cluster}
is that the outer scale of the magnetic field grows relatively quickly
after the beginning of the simulation. This is different from
direct MHD cluster simulations that have mostly
kinematic growth with a magnetic spectrum peaked on numerical dissipation
scale, e.g., \cite{Xu2012}. Note that the scale of the magnetic field plays crucial
role in cosmic ray escape times, therefore correctly estimating magnetic
outer scale is essential for models of particle acceleration in clusters~\cite[see,][]{Brunetti2007,2011MNRAS.410..127B,BXLS13,2015ApJ...800...60M}.

\section{Discussion}
Similar idea based on post-processing of hydrodynamic data was also employed in~\cite{2008Sci...320..909R}, but with substantial differences. 
The turbulence in these early calculations was not as resolved as in ours, and the growth of magnetic energy and Alfv\'en scale were not estimated from the turbulent dissipation rate and the precise estimate of $C_E$, as we did here.

One of the differences between cluster turbulence and the kind of statistically
stationary turbulence studied in \cite{B12a} was the strong variations of the cascade
rate over timescales of 1-2 dynamical timescales of the cluster. 
Our estimates of the efficiency in the case of intermittent driving from this work are roughly compatible with $C_E=0.05$ and further work with higher Re is expected to clarify whether the differences between constant and intermitted driving are significant. 
We concluded that the effects of intermittent driving could probably be ignored at the
level of precision of the $\epsilon_{turb}$ measurement.

The actual calculation for the evolution of $v_A$ and $L_B$ was started at time 4.5 Gyr. 
This artificially assumes 
that $\epsilon_{turb}$ was zero for all times earlier than $4.5$ Gyr.
However, we find that 
despite this fairly unrealistic assumption, 
the values of $v_A$ and $L_B$ quickly converge to the asymptotic values and,
as we argued above, this initial state
is quickly forgotten.
All basic properties of the cluster, such as its
mass, size and thermal energy continue to grow along with its magnetic energy and magnetic outer scale.
The detailed comparison between thermal, turbulent and magnetic energy components of the cluster
has been performed in our companion paper~\citep{miniati2015}. There it is found that 
the fraction of the thermal energy arising from the turbulent dissipation rate changes relatively little over
the cosmological time and the turbulent Mach number is also rather stable. Since the magnetic
energy is also a fraction of the accumulated turbulent dissipation rate, the plasma $\beta$
in our cluster fluctuates around a constant value $\sim 40$ for the past 10 Gyr~\citep{miniati2015}.

Our treatment of cluster turbulence with ILES, as well modeling the evolution
of the magnetic energy with the model from \cite{B12a} relies on an assumption
that the Reynolds numbers in clusters are high. For example, our
comparison of the cluster simulation and the DNS leads to an estimate of an effective
Kolmogorov (dissipation) scale for the cluster simulation of $\eta \approx 2.7$ kpc,
corresponding to an effective $Re$ around 3000. We actually expect clusters to
have higher $Re$, as briefly discussed in Section~1, due to the collective microscopic 
scattering in the high-$\beta$ ICM plasma~\citep{Schekochihin2006,LB06,Schekochihin2008,brunetti2011b}.
An important observational test to the problem of the ICM viscosity
are the measurements of Faraday rotation in AGN sources located in clusters, which allowed to probe sub-kiloparsec scales due to relatively high resolution of radio maps \citep{2011A&A...529A..13K,2008MNRAS.391..521L,Govoni2010}. The inferred magnetic spectrum in these measurements
is negative and steep, typically around Kolmogorov in the range of scales below 5 kpc and down
to the resolution limit. Such a magnetic spectrum is expected from MHD
turbulence with small dissipation scales. It would be grossly inconsistent with
magnetic spectra obtained in either kinematic dynamo models, due to their positive
spectral indexes, or with MHD models using Spitzer viscosity, which would typically
give rather shallow spectrum with index around $-1$, see, e.g. \cite{CLV02b}.
We conclude that even though it is quite obvious that magnetic Prandtl numbers
in the ICM are very high, the viscosity is not large enough to affect magnetic
spectrum above 1~kpc. Therefore, just from this observational constraint
we expect the $Re$ in clusters to be at least $10^4$ and probably much higher.
Our calculation relied on this fact and the results, grossly consistent with the current
observational properties of clusters, provide another support for the 
picture of a turbulent ICM, as opposed to an earlier view of a viscous and laminar ICM.

\def\apj{{\rm Astrophys. J.}}           
\def\apjl{{\rm ApJ }}          
\def\apjs{{\rm ApJ }}          
\def\grl{{\rm GRL }}
\def\aap{{\rm A\&A } }
\def\mnras{{\rm MNRAS } }
\def\physrep{{\rm Phys. Rep. } }               
\def\prl{{\rm Phys. Rev. Lett.}} 
\def\pre{{\rm Phys. Rev. E}} 
\def\araa{{\rm Phys. Rev. ARA\&A}} 

\bibliography{miniati,beresnyak}

\end{document}